\documentstyle[epsfig,aps,prb]{revtex}

\begin{document}
\title{Tight-binding Hamiltonians for Sr filled ruthenates$:$ application to the
gap anisotropy and Hall coefficient in Sr$_{2}$RuO$_{4}$}
\author{ I.I. Mazin, D.A. Papaconstantopoulos, and D.J. Singh}
\address{Center for Computational Materials Science, Naval
Research Laboratory,\\
Washington, DC 20375}
\maketitle
\begin{abstract}
Accurate orthogonal tight-binding Hamiltonians are constructed for
ferromagnetic SrRuO$_3$ and the layered perovskite superconductor,
Sr$_2$RuO$_4$ by fitting to all-electron full-potential local density
band structures obtained by the linearized augmented planewave method.
These Hamiltonians allow the band structure to be computed on very fine
meshes in the Brillouin zone at low cost, and additionally have
analytic band velocities, while retaining the accuracy of the full-potential
electronic structure calculations. This greatly facilitates calculation of
transport and superconducting parameters related to the fermiology.
These features are exploited to calculate the Hall coefficient and
vortex lattice geometry for Sr$_2$RuO$_4$ with fine integration meshes. We
find the lower limit for the interband order parameter anisotropy
to be compatible with the observed square geometry. We also find
that the sign reversal of the Hall coefficient can be explained in 
a conventional way if the bands are shifted by a few mRy so as to match
the experimental de Haas-van Alphen areas exactly, {\it and} the
temperature dependence of the relaxation time  is strongly dependent
on the band character.
\end{abstract}
\section*{Introduction}

The Sr-Ru-O system initially attracted interest because of 
technological applications as perovskite substrates. In the last
few years, however, the unique electronic properties of ruthenates made them
objects of interest from the point of view of fundamental solid state
physics. The root compound for this system is the pseudocubic
perovskite SrRuO$_{3}.$
This material has an orthorhombic Pbnm (GdFeO$_{3})$
structure, and is a strong ferromagnet. It is characterized by
unusual transport properties (so-called ``bad metal'' behavior),
unusually strong coupling between the spin and charge degrees of
freedom, and substantial involvement of oxygen in its magnetic properties.
Sr$_{2}$RuO$_{4}$ occurs in a body
centered tetragonal K$_{2}$NiF$_{4}$structure and is superconducting at a
temperature of 1.5 K. It is generally accepted now that the
superconductivity of
Sr$_{2}$RuO$_{4}$ is unconventional and most likely triplet.
A number of researchers expressed an opinion that ruthenates, or at least
some of them, may be strongly correlated systems close to Mott-Hubbard
transition. On the other hand, unlike common strongly correlated 
materials (3d oxides, high-$T_c$ cuprates), the physical properties 
of ruthenates are well described by conventional density functional
theory, and their unusual transport properties are likely to be 
due to strong magnetic interactions and magnetic scattering.

Density functional electronic structure calculations have been reported
for
cubic and orthorhombic SrRuO$_{3}$\cite{singh1}
and for Sr$_{2}$RuO$_{4}$%
\cite{singh2,oguchi}, using the general potential linearized 
augmented plane wave method (LAPW). An analysis of the resulting electronic
structure was given in Refs.\onlinecite{we1,we2}.
At the moment, essentially all observable properties which could be 
reliably calculated from the electronic structure results agree
well with the experiment. On the other hand,
there are several properties of
this material that cannot be reliably 
obtained directly with the LAPW method due to the high
computational cost. At the same time, there is substantial interest among
both theorists (as a starting point for many body models) and
experimentalists in having a simple, fast tight binding model which would
accurately describe the LDA electronic structure near the Fermi level\cite
{lich,dan}. Existing tight binding models\cite{we2,noce} do not reproduce 
accurately enough the details of the fermiology, e.g. the degree of nesting.
For this reason we have performed highly accurate tight-binding
parametrizations of this system by fitting to LAPW results.

\section*{Construction of the Tight Binding Hamiltonians}

For the cubic SrRuO$_{3}$ we constructed orthogonal tight-binding (TB)
Hamiltonians by fitting to the spin-polarized
band structure of Ref.\onlinecite
{singh1}. We followed a procedure similar to the one used for other
oxide perovskites\cite{jul}, based on
14 LAPW bands for each of 165 {\bf k}-points in the
irreducible Brillouin zone. Our TB Hamiltonian results in a 14$\times $14
secular equation based on the d-Ru and p-O orbitals. This Hamiltonian
contains 11 first and 2 second
 nearest neighbor Slater-Koster parameters (listed in Table \ref
{tab1}) for each spin determined by least-squares fitting to the LAPW
results. The overall rms error for all 14 bands was 14 mRy. This high level
of accuracy is demonstrated in our energy band and densities of states
figures discussed in the next section. As can be seen from Table \ref{tab1},
we have included the crystal field splitting $e_{g}-t_{2g}$ on the Ru $d$%
-states, which is 3.34 eV ($\uparrow )$ and 3.23 eV($\downarrow ).$ There is
also significant crystal field splitting $p_{\pi }-p_{\sigma }$ on the
oxygen site, 1.33 eV ($\uparrow )$ and 1.41 eV($\downarrow )$ (a similar
effect exists in manganite perovskites, cf. Ref.\onlinecite{CMR}). The exchange
splitting for Ru is relatively small: 0.57 eV ($t_{2g})$ and 0.45 eV ($%
e_{g}).$ We also have exchange splitting on the O sites (0.12 eV for $%
p_{\sigma }$ and 0.20 eV for $p_{\pi })$, in accord with the general role
of Ru-O hybridization in ruthenates, as discussed in Ref.\onlinecite{we1}. The
strongest interactions in the system are (Table \ref{tab1}) the Ru-O $%
pd\sigma $ and $pd\pi ,$ as expected. We also note that in order to obtain a
good fit we had to include the second nearest $pp\sigma $ and $pp\pi $
interactions. Another observation which can be made from Table \ref{tab1}
is that the rigid-band Stoner picture describes the physics of magnetic
splitting rather well: there is hardly any difference between the hopping
intergal in the spin up and in the spin down channels.

For Sr$_{2}$RuO$_{4}$, we followed a similar procedure to construct the
orthogonal Hamiltonian. The details are as follows. We fitted to the LAPW
results of Ref.\onlinecite{singh2} using 18 bands for each of 207 {\bf k}-points
in the irreducible Brillouin zone. The size of our Hamiltonian matrix became
27$\times $27 based on the $d-$Sr, 
$d-$Ru and $p-$O orbitals. Having in mind
applications to superconductivity theory, we aimed at having a virtually
perfect fit in the window of $\pm 1$ eV around the Fermi level. For
that inclusion of the $d-$Sr orbitals was unavoidable. This Hamiltonian
contains 30 first  and 2 second
nearest neighbor parameters. Since we used an orthogonal
TB model, the physical nonorthogonality between d-Ru and p-O orbitals (which
have bigger overlaps than other orbitals in the system) spills over into
nonzero second nearest neighbor parameters in an orthogonal Hamiltonian,
specifically Ru-O2 $pd\sigma ,$ and O1-O2 $pp\sigma $,
as well as into direct hopping integrals between Ru (the fact that
the fitted values for Ru-Ru hopping do not come from actual overlap of the
wave functions is corroborated by the fact that the $dd\delta $ parameters
have a far stronger effect on the band structure than the
$dd\sigma $ and $%
dd\pi $ ones). In our notation O1 and O2 describe the in-plane and apical
oxygens, respectively. The parameters are listed in Table \ref{tab2}.
 The overall r.m.s. error for this fit is 
$\approx 10$  mRy for all 18 bands and is less than 2
mRy for the bands crossing the Fermi level (13th to 15th),
 for which we used a higher weight in order
to reproduce the Fermi surface very accurately. 

\section*{Results}

\begin{figure}
\centerline{\epsfig{file=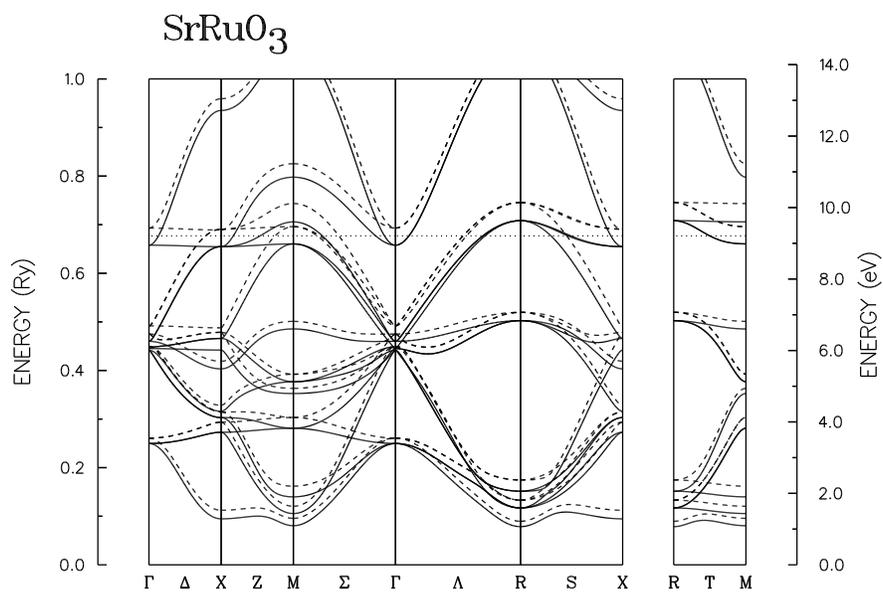,height=0.8\linewidth,angle=-90}}
\vspace{3in} \nopagebreak
\caption{TB band structure
 of cubic  SrRuO$_3$ for two spin directions.\label{sr3bands}}
\end{figure}

\begin{figure}[tbp]
\centerline{\epsfig{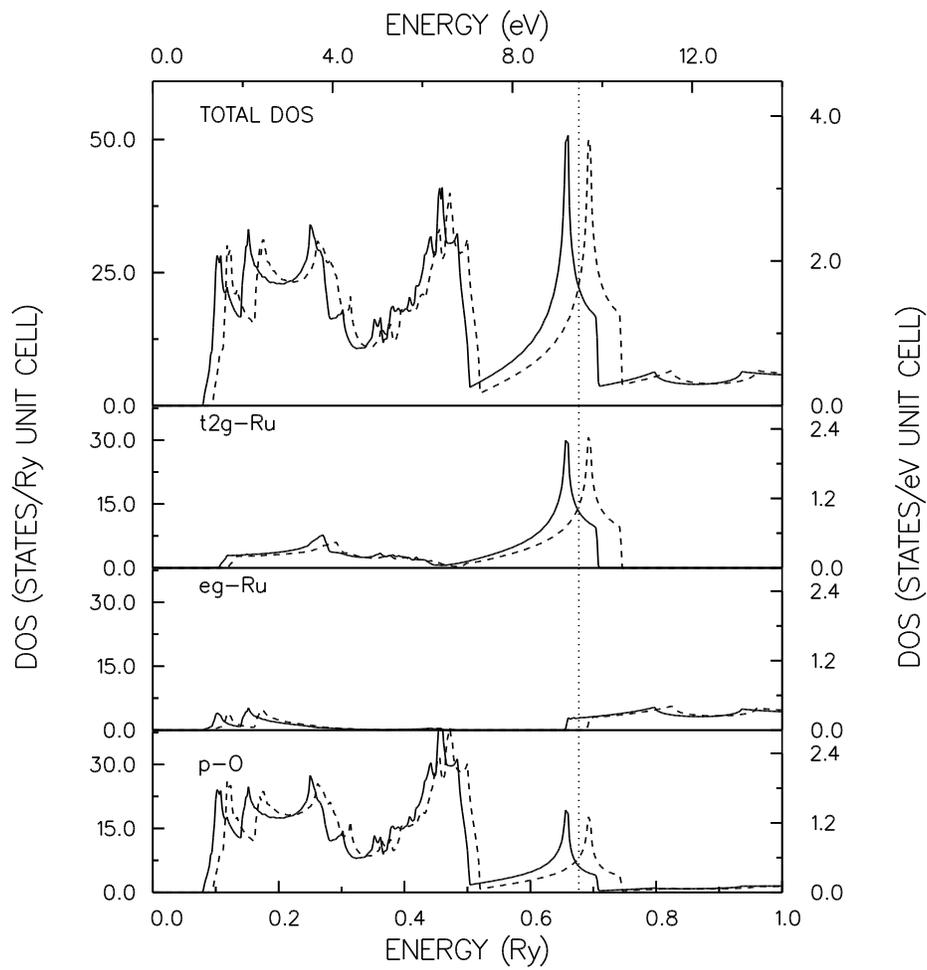}}
\vspace{.9in} \nopagebreak
 \caption{TB DOS of cubic SrRuO$_3$ for the two spin
directions.\label{sr3dos}} \end{figure}

In Fig. \ref{sr3bands} we show the TB energy bands of cubic SrRuO$_{3}$ for both spin up
and spin down. We note that the TB bands, especially near the Fermi level ($%
E_{F}$) are almost indistinguishable from the LAPW ones. In the (100)
direction we note near $E_{F}$ a flat band that is occupied for spin-up and
above E$_{F}$ for spin-down. We also see a small electron pocket for spin-up
and a gap for spin-down centered at $\Gamma $. We observe a hole pocket for
spin-down centered at M that vanishes for spin-up. In Fig. \ref{sr3dos}
we show the TB
DOS for both spins. In agreement with the LAPW results, we find a pronounced
peak just below $E_{F}$ for spin-up and just above $E_{F}$ for spin-down,
while DOS at $E_{F}$ is approximately the same for both spins.
Both these peaks are a mixture of $t_{2g}$-Ru and $p_{\pi }$-O states as can
be seen from the decomposition of the DOS, in accord with the discussion in
Ref.\onlinecite{we1}. On the other hand, the $e_g$ contribution to DOS
at the Fermi level is very small for spin up and nearly zero for the
spin down channel.

\begin{figure}[tbp]
\centerline{\epsfig{file=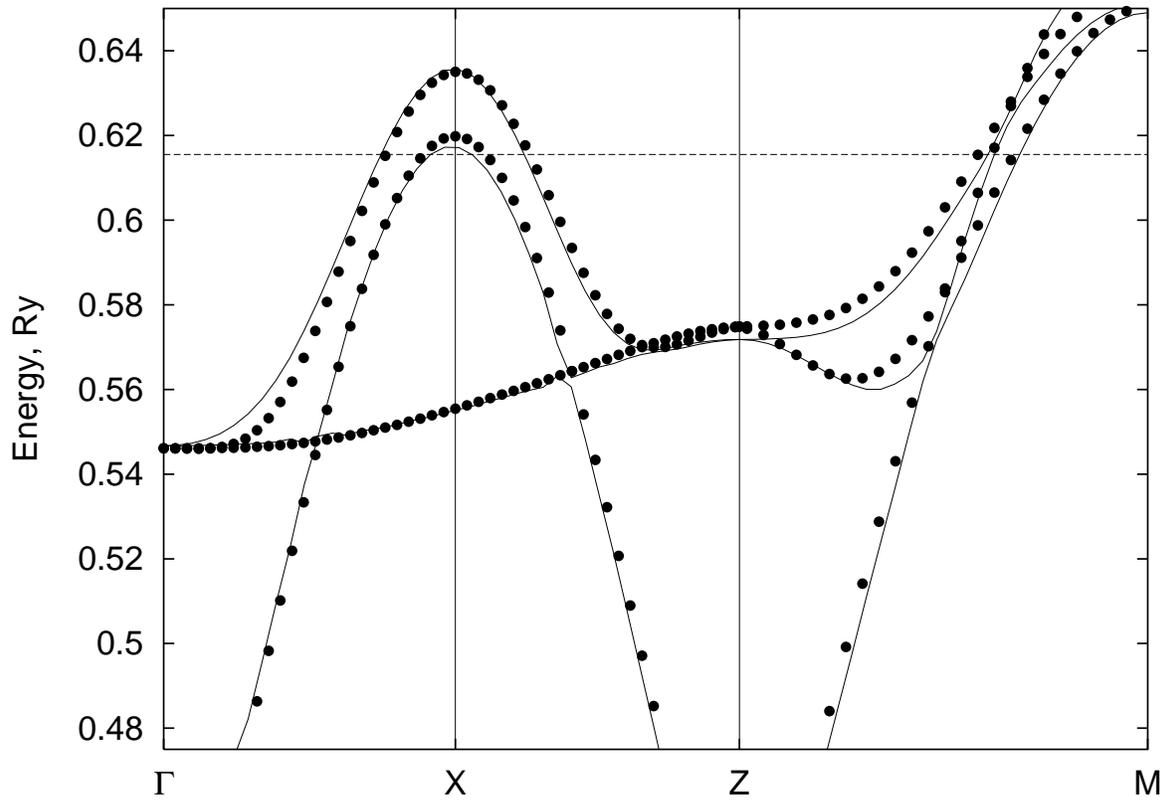,width=0.9\linewidth}}
\vspace{2in} \caption{\label{sr2bands} 
Band structure of Sr$_2$RuO$_4$ in
LAPW (circles) and in the TB fit (lines).}
\end{figure}

\begin{figure}[tbp]
\centerline{\epsfig{file=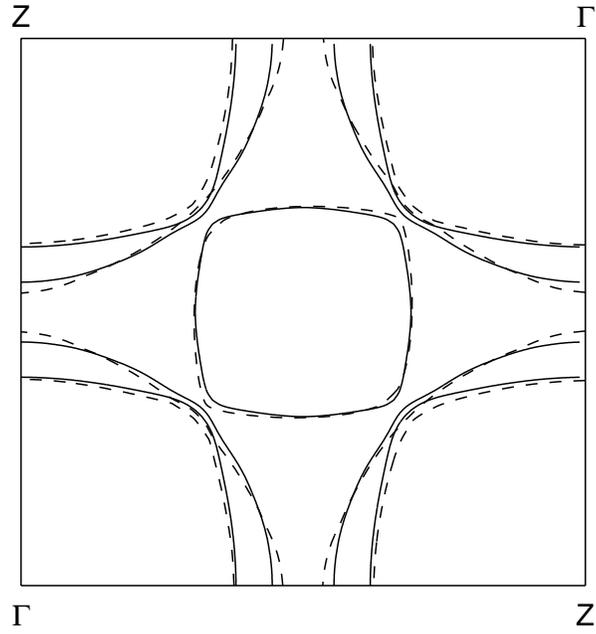,width=0.8\linewidth}}
\vspace{0.2in}  \nopagebreak
 \caption{Fermi surface of Sr$_2$RuO$_4$ in
LAPW (solid lines) and in the
 TB fit (dashed lines)\label{fs}} \end{figure}

In Fig. \ref{sr2bands}, we show a comparison of TB and LAPW energy bands for Sr$_{2}$RuO$%
_{4}$. The bands above
2 eV were not fitted. In the low bands far away from $E_{F}$ there are
minor discrepancies. However, the bands near $E_{F}$ display excellent
agreement between TB and LAPW. This also results in a virtually
perfect match of the
Fermi surfaces (Fig. \ref{fs}). The degree of the nesting of the
pseudosquare Fermi
surface sheets is reproduced very well, which is crucial for the
calculations of susceptibilities and response functions (cf. Ref.\onlinecite{AFM}%
). Also the (very weak) $k_{z}$ dispersion, which manifests itself through
small violations of the mirror symmetry in Fig.  \ref{fs}, is reproduced.

Apart from computational efficiency, TB models have another advantage with
respect to the first principles methods: electronic velocities, which are
usually calculated in the first principles methods by numerical
differentiation of electron eigenenergies, can be obtained analytically
from the TB Hamiltonian without any loss of accuracy: 
\begin{eqnarray}
{\bf v}_{{\bf k}\alpha } &=&\frac{\partial E_{{\bf k}\alpha }}
{\hbar\partial {\bf %
k}}=\left\langle {\bf k}\alpha \left| \frac{\partial H}{\partial {\bf k}}%
\right| {\bf k}\alpha \right\rangle  \label{vel} \\
\frac{\partial H}{\hbar
\partial {\bf k}} &=&\sum_{{\bf R}}i{\bf R}e^{i{\bf kR}}t_{%
{\bf R}},  \nonumber
\end{eqnarray}
where summation is performed over all neighbors for which hopping integrals
were included in the Hamiltonian. The first line follows from the
Hellman-Feynman theorem, and the second from the TB expression for $
H=\sum_{{\bf R}}e^{i{\bf kR}}t_{{\bf R}}.$ The advantage of using this
formula is particularly clear when there are many band crossings in the
energy range of interest, in which case numerical differentiation requires
an extremely fine mesh. This is even more true for the
physical quantities that depend on an
electronic mass (second derivative of the eigenenergy), like Hall
coefficient. Since the error of numerical differentiation increases rapidly
with the order of the calculated derivative, eliminating the second
differentiation by using the Hellman-Feynman velocities is particularly
helpful in this case. To illustrate
this approach, we computed a quantity that defines
the vortex lattice geometry in superconducting Sr$_{2}$RuO$_{4}$\cite{dan}
and depends in a non-trivial way
on the electronic velocities.
Agterberg\cite{dan} showed that this  geometry is
directly related to the symmetry of the order parameter. For an isotropic
triplet $p$-wave pairing, one expects a triangular vortex lattice. For a $d$
wave pairing a square lattice appears, oriented either along (100) or along
(110) directions. It was observed experimentally\cite{riseman} that in Sr$%
_{2}$RuO$_{4}$ Abrikosov vortices form a square lattice; Agterberg\cite{dan}
performed calculations for a model of a 2D triplet superconductor, where the
angular anisotropy comes entirely from the fermiology. He derived a
criterion that controls  the geometry of the vortex lattice: 
\begin{equation}
\nu =\frac{2\left\langle \delta v^{2}\right\rangle ^{2}-\left\langle
v^{2}\right\rangle ^{2}}{\left\langle v^{2}\right\rangle ^{2}}=2\frac{%
\left\langle v_{x}^{2}-v_{y}^{2}\right\rangle ^{2}}{\left\langle
v_{x}^{2}+v_{y}^{2}\right\rangle ^{2}}-1.  \label{nu}
\end{equation}
The absolute value of this quantity determines whether the vortex lattice
will be triangular or square (for $\nu >0.014$ it is always square), while
its sign defines the orientation of the lattice when it is square. If $\nu
<0 $ it is aligned with the crystal lattice, as it is observed in the
experiment.

Agterberg pointed out that when order parameters are different for
the three bands crossing the Fermi level, the actual vortex lattice will be
defined by the sign of the weighted $\nu $ parameter, which in turn
depends on the order parameters $\Delta$ in all three bands, 
\begin{equation}
\nu _{eff}=(\nu _{\alpha }\left\langle v^{2}\right\rangle _{\alpha
}^{2}\Delta _{\alpha }^{2}+\nu _{\beta }\left\langle v^{2}\right\rangle
_{\beta }^{2}\Delta _{\beta }^{2}+\nu _{\gamma }\left\langle
v^{2}\right\rangle _{\gamma }^{2}\Delta _{\gamma }^{2})/(\left\langle
v^{2}\right\rangle _{\alpha }^{2}\Delta _{\alpha }^{2}+\left\langle
v^{2}\right\rangle _{\beta }^{2}\Delta _{\beta }^{2}+\left\langle
v^{2}\right\rangle _{\gamma }^{2}\Delta _{\gamma }^{2}).  \label{eff}
\end{equation}
He estimated, using a 3-band TB model of Ref. \onlinecite{we2}, that $\nu _{\alpha
}\approx \nu _{\beta }\approx 0.6,$ while $\nu _{\gamma }\approx -0.6.$ This 
calculation is to be taken with a grain of salt, though, as the model used
for the band structure is rather approximate. 
However, if indeed $\nu _{\alpha }\approx \nu _{\beta }\approx -\nu _{\gamma
}>0,$ even a modest interband anisotropy of the order parameter such as that
computed in Refs.\onlinecite{we2,AFM} could ensure agreement with the experiment.
Of course, since Eqs. \ref{nu},\ref{eff} involve complicated averages over
the Fermi surface, the result is very sensitive to the accuracy of the
Brillouin zone integration.

\begin{figure}[tbp]
\centerline{\epsfig{file=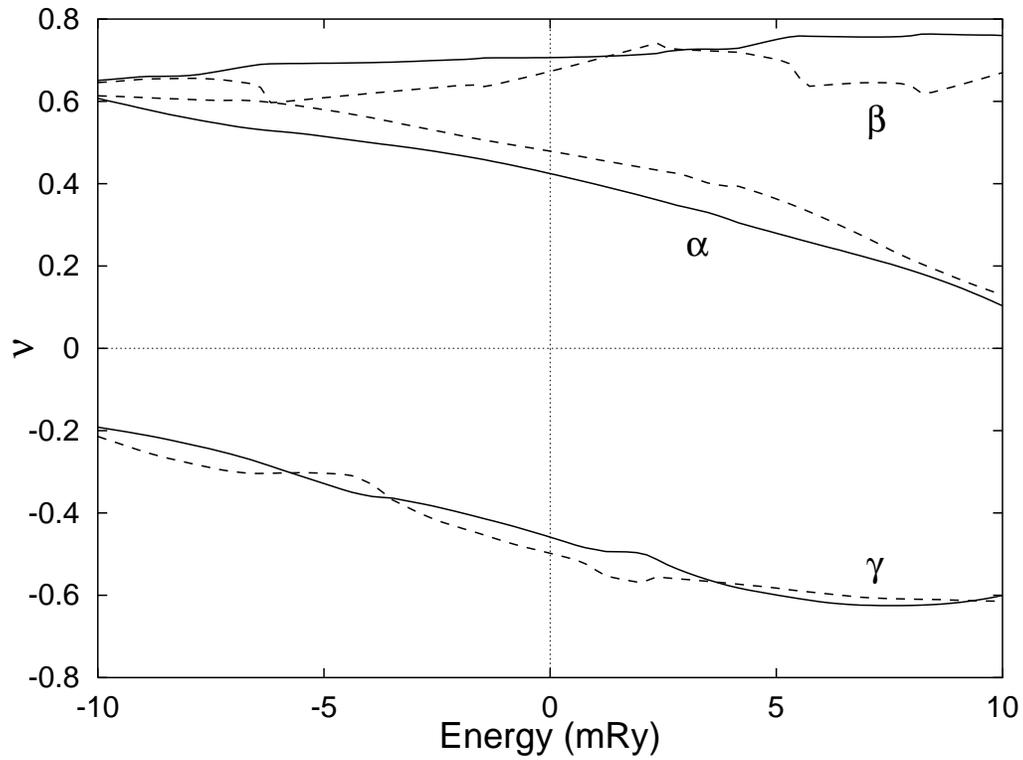,width=0.8\linewidth}}
\vspace{2in} \caption{Anisotropy parameter $\nu$ (see text)
as the function of the distance to the Fermi level. Solid lines
are results of the integration with exact TB velocities, broken
lines use velocities from tetrahedron linear interpolation.
In both cases 30$\times 30 \times 10$ mesh was used.\label{nufig}}
\end{figure}

In Fig.\ref{nufig} we show the results of our calculation of $\nu$,
using the
present TB model, Eq.\ref{vel}, and the tetrahedron integration technique
with the 1507 irreducible {\bf k}-points in the Brillouin zone. First, we
observe that even with such a fine mesh and relatively simple Fermi surface
the numerical differentiation of eigenenergies produces noisy curves,
compared to the analytical differentiation. Second, we find that there is
little dependence of $\nu $ on the Fermi energy in the range $\pm $150 meV,
although the van Hove singularity is located only 70 meV above the Fermi
energy. Third, we see that the calculated values of $\nu $ ($\nu _{\alpha
}=0.42,$ $\nu _{\beta }=0.70,$ $\nu _{\gamma }=-0.45)$ are not too far from
Agterberg's estimates. Fourth, we can use Eq. \ref{nu} to estimate the lower
limit on the ratio $\Delta _{\gamma }/\Delta _{\alpha },$ assuming $\Delta
_{\alpha }=\Delta _{\beta }.$ In order to reproduce the experimental
observation that $\nu _{eff}<0,$ and using our calculated ratios $%
\left\langle v^{2}\right\rangle _{\alpha }:\left\langle v^{2}\right\rangle
_{\beta }:\left\langle v^{2}\right\rangle _{\gamma }=1.0:0.55:1.0,$ we find
that $\Delta _{\gamma }/\Delta _{\alpha }>1.2,$ which is to be compared, for
instance, with $\Delta _{\gamma }/\Delta _{\alpha }\approx $2.2 in Ref. \onlinecite
{we2}.

Another interesting Fermi surface property, which requires highly accurate
integration over the Brillouin zone, is the
Hall coefficient. It depends on the 
{\it second} derivatives of the eigenvalues (see Appendix), which makes
numerical differentiation undesirable. We used our TB fit to calculate the
Hall coefficient both in diffusive (Eq.\ref{Rd}) and ballistic(Eq.\ref{Rb})
regimes. Experiments show \cite{andy} that the Hall coefficient is negative
(electron-like) at $T\rightarrow 0,$ $R=-1.15\times 10^{-10}$ m$^{3}/$C, but
grows rapidly and changes sign at $T\approx 35$ K, and stays close to zero
at the higher temperatures. It was suggested that this is a manifestation of
the different temperature dependence of the relaxation time in different
bands\cite{andy}. Alternatively, one may think that at $T\sim 35$ K a
crossover between the Eq.\ref{Rb} and Eq.\ref{Rd} takes place.

We checked these hypotheses by calculating the conductivities $\sigma _{xy}$
and $\sigma _{xx}$ (see Appendix)
for all three Fermi surface sheets. In both cases they
are calculated up to a common constant factor (constant relaxation time $%
\tau $ or constant mean free path $\lambda $). The results were, in atomic
units, as given in Table \ref{hall}. First of all, we observe
that the difference
between the two approximations is very small. Second, we see that the net
Hall conductivity originating
 from the $xz/yz$ sheets of the Fermi surface ($
\alpha $ and $\beta $, in standard notations) is electron-like. Thus the
Mackenzie $et$ $al$ explanation of the sign change at $T\sim 35$ K requires
a substantial difference of the temperature dependence of $\tau _{\alpha }$
and $\tau _{\beta }$ (not only of $\tau _{\alpha ,\beta }$ and $\tau
_{\gamma }),$ which is physically hard to support, considering the 
origin of these bands.

However, as noted in a previous paper\cite{we2}, to fit the
experimental de Haas-van Alphen extremal area exactly, one has to
shift the LDA
bands by a few mRy up ($\alpha )$ or down ($\beta ,\gamma ).$ In order to
investigate this, we applied shifts of 5, -4, and -1.5 mRy to the bands $%
\alpha ,$ $\beta ,$ and $\gamma ,$ respectively\cite{note} This procedure
reduces of the energy distance between the van Hove singularity and the
Fermi level from $\approx 60$ to $\approx 20$ meV, which makes the
difference between the ballistic and the diffusive $\sigma _{xy}$ more
pronounced. The results are shown in Table \ref{shift}. One immediately observes that
now the net Hall conductivity in the bands $\alpha $ and $\beta $ is {\it %
positive.} Thus if the mean free path in those two bands changes with
temperature slower than that for the $\gamma $ band (to some extent, in
spirit of the ``orbital-dependent'' conjecture of Agterberg {\it et al}\cite
{ASR}), one may, in principle, expect a sign change of the net Hall
coefficient. In fact, one needs to assume that the ratio $\lambda _{\alpha
,\beta }/\lambda _{\gamma }$ increases from $\approx 1$ at $T=0$ to $\approx
4$ at $T\approx 40$ K to explain the observed sign change. This would be
unusual, but certainly not impossible.

\section*{Conclusions}
To conclude, we presented highly accurate tight binding fits for 
cubic SrRuO$_3$ and tetragonal Sr$_2$RuO$_4$ perovskites. 
Using nearest neighbor and a few selected next nearest neighbor
hopping parameters
 we were able to reproduce the first-principles band structure
of Sr$_2$RuO$_4$ in the physically relevant energy range near the 
Fermi level with the accuracy of a few mRy. This allows for 
fast generation of the electronic eigenenergies at a superfine 
mesh in the Brillouin zone, and fast and virtually exact calculations
of the electronic velocities at the same mesh.
This powerful technique provides a possibility of investigating
physical properties with high sensitivity to the details of the
band structure and Fermi surface topology.

We report two application of the present fit to actual physical problems:
First, we calculated the anisotropy parameter $\nu$ relevant for the
superconducting vortex lattice geometry, and derived from that 
the lower bound for the interband anisotropy of the order parameter:
to ensure compatibility with the experiment, the order parameter 
in the $xy$ band should be at least  20\% larger than in the $yz/zx$
bands. Second, we calculated Hall conductivity for the three bands 
crossing the Fermi level, both in the constant-relaxation-time and in the
constant-mean-free-path regimes. Having these numbers  we
tested the two most natural hypotheses for the temperature-induced
sign reversal of the Hall conductivity: crossover between the two 
above-mentioned regimes and different temperature dependence of
the $xy-$ and $yz/zx-$carrier mobility.  We can confidently 
rule out the former hypothesis.  The latter hypothesis cannot
be definitely ruled out; for the Fermi surface adjusted to fit
the de Haas - van Alphen cross-section the desired sign reversal
can be obtained in the constant-mean-free-path regimes, if  the mean
free path for the $xy-$carriers changes between O K and 40 K at least
four times faster than that for the $yz/zx-$carriers. A possible reason
for such a behavior would be substantially stronger electron-paramagnon
coupling for the $xy-$electrons.
\acknowledgements
We are grateful to D. Agterberg for useful discussions. This
work was supported by the Office of Naval Research.

\appendix
\section*{Derivation of the Hall formulas}

It is convenient to use the 2D expression given by Ong\cite{ong}: 
\begin{eqnarray}
\sigma _{xy} &=&2\frac{e^{2}B}{\hbar \Omega }\sum_{{\bf k}i}\delta (E_{{\bf k%
}i}-E_{F})v_{y}({\bf k)\tau }_{{\bf k}i}\left[ v_{y}({\bf k)}\frac{\partial 
}{\partial k_{x}}-v_{x}({\bf k)}\frac{\partial }{\partial k_{y}}\right]
v_{x}({\bf k)\tau }_{{\bf k}i} \\
\sigma _{xx} &=&2\frac{e^{2}}{\hbar \Omega }\sum_{{\bf k}i}\delta (E_{{\bf k}%
i}-E_{F})v_{x}^{2}({\bf k)\tau }_{{\bf k}i},
\end{eqnarray}
which defines the Hall coefficient, $R=-\sigma _{xy}/B\sigma _{xx}^{2}.$
Here $\Omega $ is the unit cell volume, and summation is over all bands and
all states in the Brillouin zone. In the constant-relaxation-time
approximation this reduces to the following expression: 
\begin{equation}
R_{d}=-\sum_{{\bf k}i}\delta (E_{{\bf k}i}-E_{F})(v_{x}^{2}\mu
_{yy}-v_{x}v_{y}\mu _{xy})/2\left[ \sum_{{\bf k}i}\delta (E_{{\bf k}%
i}-E_{F})v_{x}^{2}\right] ^{2},  \label{Rd}
\end{equation}
where $\mu _{xy}({\bf k)}=\partial ^{2}E_{{\bf k}}/\partial k_{x}\partial
k_{y}$ etc. As noticed by Mackenzie {\it et al}\cite{andy}, at very low
temperature a more appropriate approximation is where the mean free path, $v_{%
{\bf k}}$ ${\bf \tau }_{{\bf k}}$ is a constant, in which case we have
instead 
\begin{equation}
R_{b}=\sum_{{\bf k}i}\delta (E_{{\bf k}i}-E_{F})(v_{x}^{2}\mu
_{yy}-v_{x}v_{y}\mu _{xy})v^{-2}/2\left[ \sum_{{\bf k}i}\delta (E_{{\bf k}%
i}-E_{F})v_{x}\right] ^{2}.  \label{Rb}
\end{equation}

\eject
\vspace {-1in}

\eject

\eject

\eject
\eject

\eject
\begin{table}
\caption{
Slater-Koster parameters for the cubic SrRuO$_{3}$ (mRy).}
\label{tab1}
\begin{tabular}{c|cc|cc|ccc|cc|cc|cc}
spin& \multicolumn{2}{c|}{Ru onsite} & \multicolumn{2}{c|}{O onsite}&\multicolumn{3}{c|}{Ru-Ru}
 & \multicolumn{2}{c|}{O-O} & \multicolumn{2}{c|}{Ru-O}&\multicolumn{2}{c}{O-O}\\
& $t_{2g}$ &  $e_{g}$ &  $p_\sigma$ &  $p_\pi$ & $dd\sigma $ & $dd\pi $ & $%
dd\delta $ & $pp\sigma $ & $pp\pi $ & $pd\sigma $ & $pd\pi $ & $pp\sigma
^{a} $ & $pp\pi ^{a}$ \\  \tableline
$\uparrow $ & 493 & 738 & 307 & 405 & -28 & -12 & 1 & 43 & -11 & 168 & -94 & 
14 & -4 \\ 
$\downarrow $ & 534 & 772 & 316 & 420 & -27 & -11 & 1 & 43 & -11 & 165 & -97
& 14 & -5
\end{tabular}

$^{a}$second nearest neighbors
\end{table}

\begin{table}[tbp]
\caption{
\label{tab2}
Slater-Koster parameters for Sr$_{2}$RuO$_{4}$ (mRy).}
\begin{tabular}
{ccc|c|cc|ccc|ccc|cc|cc|ccc}
\multicolumn{3}{c|}{Ru onsite}& O1 onsite &\multicolumn{2}{c|}{O2 onsite}&\multicolumn{3}{c|}{Sr onsite}&\multicolumn{3}{c|}{Ru-Ru} &
 \multicolumn{2}{c|}{O1-O1}& \multicolumn{2}{c|}{O2-O2}
 & \multicolumn{3}{c}{O1-O2}\\
\tableline
$xy$ & $yz,zx$ & $e_{g}$ & $p$ & $z$ &  $x,y$ &  $xy$ &  $yz,zx$ & $e_{g}$& $dd\sigma $&$dd\pi $ &$dd\delta $ &  $pp\sigma $ 
&  $pp\pi $ &  $pp\sigma $&  $pp\pi $ &  $pp\sigma$&$pp\pi$&  $pp\sigma
 ^{a}$
 \\ 
487    & 516         & 659        & 338    &    392 &      436 & 1032    & 1318       & 1874      &-7 &-3&-6   & 37
& -7           &              7 & 4& 44  &-18&-7
\end{tabular}
\begin{tabular}
{c|ccc|cc|cc|cc|ccc}
Sr-Sr& \multicolumn{3}{c|}{Sr-Ru} & \multicolumn{2}{c|}{O1-Sr} & \multicolumn{2}{c|}{O1-Ru} & \multicolumn{2}{c|}{O2-Sr} & 
\multicolumn{3}{c}{O2-Ru} \\ 
 \tableline
$dd\pi $ &$dd\sigma $ & $dd\pi $ & $dd\delta $ & $pd\sigma $ & 
$pd\pi $ & $pd\sigma $ & $pd\pi $ & $pd\sigma $ & 
$pd\pi $ & $pd\sigma $
&$pd\pi $ & $pd\sigma ^{a}$  \\ 
39& -80 & -7 & -38 & -47 & 110 & 190 & 96 & 61 & -58 & 160
&68 & 17 
\end{tabular}

$^{a}$second nearest neighbors
\end{table}

\begin{table}[tbp]
\caption{Conductivities and Hall coefficients in atomic units.
The volume of the unit \label{hall}
cell $\Omega =$95.4 \AA $^{3},$ $\tau $ and $\lambda $ are the relaxation
time and the mean free path, respectively. Atomic units for $R$ are such
that for a parabolic band $R=1/n,$ where $n$ as given in the first columns
is the total number of electrons ($\alpha ,$ $\gamma )$ or holes ($\beta )$
in the corresponding bands. }
\begin{tabular}{ccccccccc}
band & $n$ & $\sigma _{xx}^{d}/\tau \Omega $ & $\sigma _{xx}^{b}/\lambda
\Omega $ & $\sigma _{xy}^{d}/\tau \Omega $ & $\sigma _{xy}^{b}/\lambda
\Omega $ & $R_{d}\Omega $ & $R_{b}\Omega $ & $R_{\exp }\Omega (T=0)$ \\ 
\tableline
$\alpha $  & -1.02 & 2.0 & 6.0 & -4.0 & -36 & -0.95 & -0.98 &  \\ 
$\beta $  & 0.28 & 1.1 & 3.2 & 3.8 & 32 & 3.2 & 3.2 &  \\ 
$\gamma $  & -1.26 & 2.0 & 6.6 & -2.8 & -30 & -0.65 & -0.68 &  \\ 
total & -2.00 & 5.1 & 15.2 & -3.0 & -34 & -0.12 & -0.15 & -0.20
\end{tabular}
\end{table}

\begin{table}[tbp]
\caption{The same as Table \ref{shift}, but for the bands shifted to match
de Haas van Alphen cross section areas.\label{shift}}
\begin{tabular}{ccccccccc}
band & $n$ & $\sigma _{xx}^{d}/\tau \Omega $ & $\sigma _{xx}^{b}/\lambda
\Omega $ & $\sigma _{xy}^{d}/\tau \Omega $ & $\sigma _{xy}^{b}/\lambda
\Omega $ & $R_{d}\Omega $ & $R_{b}\Omega $ & $R_{\exp }\Omega (T=0)$ \\ 
\tableline
$\alpha $ 3 & -0.92 & 2.0 & 5.7 & -4.0 & -35 & -1.03 & -1.05 &  \\ 
$\beta $ 1 & 0.25 & 1.0 & 3.0 & 3.4 & 37 & 3.4 & 4.17 &  \\ 
$\gamma $ 2 & -1.35 & 1.9 & 6.7 & -1.8 & -29 & -0.52 & -0.64 &  \\ 
total & -2.02 & 4.9 & 15.4 & -3.4 & -31 & -0.14 & -0.13 & -.20
\end{tabular}
\end{table}

\begin{references}
\bibitem{note}  The last number differs from the one in Ref.\onlinecite{we2} (3
mRy) because of the difference between the TB fit and the LAPW results.
\end{references}
\end{document}